\documentclass[pre,reprint,superscriptaddress]{revtex4-1}

\usepackage{amsmath}
\usepackage{amssymb}
\usepackage{graphicx}

\usepackage{textcomp}

\begin{document}
\title{Local clustering coefficient based on three-way partial correlations in climate networks as a new marker of tropical cyclone}

\author{Mikhail Krivonosov}
\affiliation{Lobachevsky State University of Nizhny Novgorod, Nizhny Novgorod, 
	Russia}
\author{Olga Vershinina}
\affiliation{Lobachevsky State University of Nizhny Novgorod, Nizhny Novgorod, 
	Russia}
\author{Anna Pirova}
\affiliation{Lobachevsky State University of Nizhny Novgorod, Nizhny Novgorod, 
	Russia}
\author{Shraddha Gupta}
\affiliation{Potsdam Institute for Climate Impact Research, PO Box 601203, 14412 Potsdam, Germany}
\affiliation{Department of Physics, Humboldt University, Berlin, Germany}
\author{Oleg Kanakov}
\affiliation{Lobachevsky State University of Nizhny Novgorod, Nizhny Novgorod, 
	Russia}
\author{J\"urgen Kurths}
\affiliation{Lobachevsky State University of Nizhny Novgorod, Nizhny Novgorod, Russia}
\affiliation{Potsdam Institute for Climate Impact Research, PO Box 601203, 14412 Potsdam, Germany}
\affiliation{Department of Physics, Humboldt University, Berlin, Germany}
		
\begin{abstract}
We introduce a new network marker for climate network analysis. It is based upon an available special definition of local clustering coefficient for weighted correlation networks, which was previously introduced in the neuroscience context and aimed at compensating for uninformative correlations caused by indirect interactions. We modify this definition further by replacing Pearson's pairwise correlation coefficients and Pearson's three-way partial correlation coefficients by the respective Kendall's rank correlations. This reduces statistical sample size requirements to compute the correlations, which translates into the possibility of using shorter time windows and hence into shorter response time of the real-time climate network analysis. We compare this proposed network marker to the conventional local clustering coefficient based on unweighted networks obtained by thresholding the correlation matrix. We show several examples where the new marker is found to be better associated to tropical cyclones than the unweighted local clustering coefficient.
\end{abstract}

\maketitle
	

\section{Introduction}
Network analysis is recognized as a powerful tool in climate science nowadays, in essence being a specialization of the general correlation networks approach \cite{langfelder2008wgcna}, which is also well established in systems biology \cite{bruggeman2007nature, van2007art, friedman2012inferring, rosato2018correlation} and neuroscience \cite{fransson2008precuneus, bassett2017network}. Correlation networks in climate science are constructed based upon cross-correlations in multivariate time series of chosen climate variables between nodes of a spatial grid (e.g. a geographic coordinate grid on Earth surface). Such networks are then used to compute various graph metrics, which in turn become input features for subsequent analysis to identify patterns of climate dynamics \cite{tsonis2006networks, donges2009complex, gozolchiani2011emergence, dijkstra2019networks}.

It was suggested in \cite{feng2016climatelearn} and in follow-up publications (see \cite{dijkstra2019application} for review) to use climate network metrics as input features to machine learning algorithms, and this approach was successfully applied to predicting El Ni\~no phases. From the machine learning perspective, network analysis can be viewed as a preprocessing stage to reduce the dimensionality of input data. Without such preprocessing, the input data at any specified moment of observation would include all data in a time section of specific duration (sliding time window preceding the current moment) of the multivariate time series. The dimensionality of such data is the product of the number of spatial grid points and the number of time samples within the window. Network analysis reduces this dataset to a few graph metrics. In case of local metrics, which are computed for a specific geographic location associated to a network node (e.g. node degree or local clustering coefficient), machine learning methods can be applied also locally, with the set of local network metrics at a specified geographic point taken as input features, while the metrics themselves are computed, generally speaking, based on full multivariate input data. Thus, network analysis in the context of machine learning allows to cut down the dimensionality of the learning problem by orders of magnitude, while still retaining the dependence of the outcome upon all available data.

It is essential for the efficiency of this approach that as much as possible of information contained in the initial data be retained in the feature set produced by network analysis. A common practice in the literature on climate networks is to construct an unweighted graph based upon the correlation matrix of the input data by simple thresholding: network nodes are associated to the spatial grid nodes of the input data, and a link between two specific nodes is assumed to exist if the corresponding component of the correlation matrix (i.e. the correlation coefficient of the corresponding variables) exceeds a chosen threshold. The threshold value is a free parameter of the method, and generally allows tuning to maximize performance (in whatever quantitative sense), but it is also common to fix the threshold value by specifying the edge density (defined as the fraction which linked pairs of nodes constitute among the total number of node pairs), e.g. at 5\%, without any further optimization. Regardless of whether the threshold value is optimized or not, thresholding inevitably leads to information loss.

A known way to mitigate the information loss introduced by thresholding is to construct and analyze an ensemble of several unweighted graphs at once, by using a series of threshold values, which has been successfully applied to genetic network analysis \cite{bockmayr2013new}. It is also possible to eliminate the thresholding operation at all, by analyzing a full weighted graph whose edge weights are determined by (in the simplest case, taken equal to) the respective pairwise correlation coefficients. The transition from unweighted to weighted networks calls for the respective extension of graph metrics definitions. For many metrics such extensions are available \cite{Newman2016}, but may be not unique, thus giving rise to an additional problem of choosing the best weighted-network modification of a particular graph metric. An important consideration to guide this choice is that the networks of interest are correlation networks.

This approach was successfully followed in the neuroscience context \cite{frontiers2018}, where an improved formulation of local clustering coefficient (LCC) specially focused on weighted correlation networks was proposed and demonstrated to outperform the conventional thresholding approach in terms of an illustrative neuroscience problem (revealing the age dependence of human brain network structure based on functional magnetic resonance data). The key distinctive feature of the special definition of LCC for weighted correlation networks \cite{frontiers2018} consists in accounting for three-way partial correlations in order to compensate for spurious correlations caused by indirect interactions, which otherwise disguise the true interaction structure in the correlation matrix.

No literature is currently available to implement this approach in climate science. The present study aims at filling this gap by adapting the correlations-focused weighted-network LCC definition from \cite{frontiers2018} to climate network analysis. We further modify the LCC definition of \cite{frontiers2018} by switching from (parametric) Pearson's to (non-parametric) Kendall's correlations (which moves from capturing linear to more general monotonic dependencies between variables and imposes weaker requirements to sample size as compared to both Pearson's and Spearman's correlation \cite{bonett2000sample}) and relate the LCC of mean sea level pressure correlation networks to available data on tropical cyclones. We show that the LCC modification based on three-way partial correlations outperforms the unweighted thresholded network LCC as a marker of tropical cyclone.

\section{Methods}
\subsection{Cimate data}
As a source of multivariate time series for constructing climate correlation networks we use ERA5 reanalysis data \cite{hersbach2020era5}, which is essentially a model-based interpolation of available climatic observational data to a regular grid over geographic coordinates and time. We use time series of mean sea level pressure (MSLP) on a coordinate grid with step of 0.75\textdegree{} over latitude and longitude, and time grid with step of 3 hours. The data are taken for the sea surface within the range 5\textdegree N to 30\textdegree N latitude and 50\textdegree E to 100\textdegree E longitude (northern part of the Indian ocean), the time range covers 38 years from 1982 to 2019.

We compute pairwise correlations of MSLP anomalies between the nodes of the geographic coordinate grid. Anomaly is defined as the deviation of the ERA5 reanalysis data from local daily climate normals, which in turn are computed by averaging the observable (here, MSLP) at the specific location (grid node) over all daily time samples for the particular date of year and over all years of observation (for simplicity, data for 29 February of leap years are discarded).

We relate our findings to the Best Track data \cite{mohapatra2012best} on tropical cyclones in the region, which are obtained from the Regional Specialized Meteorological Centre for Tropical Cyclones Over North Indian Ocean (India) and contain information on the position and strength of all registered cyclones in the region over time.

\subsection{Correlation analysis}\label{sec:kendall}
To quantify correlation, we use the Kendall's rank correlation coefficient $\tau$ \cite{kendall1938new} (more precisely, the Goodman-Kruskal version thereof, see below), which is defined for a joint bivariate sample (taken for definiteness on a grid of observation times $\{t_i\}$) of two random variates $(x_1(t_i), x_2(t_i))$ as a normalized difference between the counts of concordant and discordant pairs of bivariate observations\footnote{A pair of bivariate observations $(x_1(t_i), x_2(t_i))$ and $(x_1(t_j), x_2(t_j))$ is called concordant (discordant), if the product $(x_1(t_i)-x_1(t_j)) (x_2(t_i)-x_2(t_j))$ is positive (negative), implying that this particular pair of observations shows an increasing (decreasing) dependence between $x_1$ and $x_2$.}. Normalization is meant to ensure that $\tau\in[-1,1]$, and in particular, $\tau=1$ ($\tau=-1$) implies a deterministic increasing (decreasing) functional dependence between $x_1$ and $x_2$. Different formulations for the normalizing denominator are available in the literature, depending on the chosen way to account for ties (i.e. equality cases with $x_k(t_i)=x_k(t_j)$), which is still a matter of research \cite{amerise2015correction}. As long as ties are negligible in our problem due to the continuous nature of climate variables, the method of their resolution is not actually significant. For definiteness, we opt to discard tied pairs if such occur, and define the normalizing denominator as the total of concordant and discordant pairs, as suggested in \cite{adler1957modification}, \cite[Eq.~(15.2) and below therein]{kruskal1958ordinal}, the resultant quantity also known as the Goodman-Kruskal gamma coefficient \cite[Eq.~(2.39) therein]{gibbons2003nonparametric}, named after authors who applied it to variates taking on finite sets of values \cite[Eq.~(21) therein]{goodman1954measures}.

Our choice of the Kendall's rank correlation coefficient over other available measures of association is due to the following considerations: (i) it imposes weaker requirements on sample size compared to both Pearson's and Spearman's correlation coefficients \cite{bonett2000sample}, which translates into shorter length of the sliding time window, and hence better time resolution; (ii) like any rank-order statistics (including Spearman's correlation), it is invariant to monotone nonlinear transformations of variables \cite[Sec.~5.5]{gibbons2003nonparametric}, which is not the case with Pearson's correlation being only invariant to linear transformations (essentially, Kendall's and Spearman's correlation assess the proximity of a bivariate sample to an arbitrary monotone dependence between variates, instead of proximity to linear dependence as in Pearson's correlation); (iii) it is more robust to outliers (large-amplitude noise) than both Pearson's and Spearman's correlation coefficients \cite{xu2013comparative}.

Correlation coefficient attributed to a particular moment of time is computed over a time window of 10 days preceding the given moment. We developed an optimized online algorithm to compute Kendall's rank correlation coefficient over a sliding time window, which benefits from the reuse of computation results obtained at the previous position of the sliding window.

\subsection{Network metrics}
We consider both weighted and unweighted networks, the former represented by the full unchanged matrix of Kendall's pairwise rank correlation coefficients $(\tau_{ij})$, and the latter constructed by thresholding the correlation matrix: a pair of nodes are considered connected when their correlation coefficient exceeds a threshold, which in turn is chosen so that the fraction of connected nodes among the total number of node pairs (edge density) is a specified quantity (taken equal to 5\% for definiteness).

Local clustering coefficient (LCC) for a given node $i$ on an unweighted graph is defined as the ratio of the number of closed triangles (connected triplets of nodes) containing the node $i$ to the total number of node triplets constructed from this node and its adjacent nodes \cite{watts1998collective}. LCC may be expressed in terms of the adjacency matrix $a_{ij}$ of an unweighted graph (by definition, $a_{ij}=1$ if a link is present between the nodes $i$ and $j$, and $a_{ij}=0$ otherwise) as
\begin{equation}\label{eq:LCC}
LCC_i^{\text{unw}} = \frac{\sum_{1\le j < l \le N, (j,l\neq i)} a_{ij} \cdot a_{il} \cdot a_{jl}}
  {\sum_{1\le j < l \le N (j,l\neq i)} a_{ij} \cdot a_{il}},
\end{equation}
where $N$ is the total number of nodes in the network, and indices $i$, $j$, $l$ enumerate the nodes. Note that the denominator in \eqref{eq:LCC} is essentially the number of unordered pairs (2-combinations) among the nodes adjacent to $i$ and can be equivalently expressed as $k_i (k_i-1)/2$, where $k_i$ is the degree of the node $i$ (the number of the nodes adjacent to $i$). We show the formulation \eqref{eq:LCC} due to its straightforward correspondence with the weighted-network extension of the LCC definition (see Eq.~\eqref{eq:LCCw} below).

Several extensions of the LCC definition to account for weighted networks are available in the literature \cite{frontiers2018, saramaki2007generalizations, rubinov2010complex, rubinov2011weight, wang2017comparison}. The version of \cite{frontiers2018} was specifically focused on correlation networks, as opposed to other LCC definitions adapted to weighted graphs in general. The idea behind this definition is to compensate for the impact that indirect interaction paths (e.g. node $j$ interacting with node $i$, which in turn interacts with node $l$) inevitably have upon the correlation matrix. Namely, the nodes $j$ and $l$ in the example above are expected to exhibit some correlation even in the absence of actual direct interaction between them, implying that such indirect correlations disguise the true interaction structure in the correlation matrix. In order to minimize the influence of such indirect correlations on the resultant LCC value, the definition of LCC in \cite{frontiers2018} makes use of the three-way partial Pearson's product-moment correlation coefficient $\rho^{\text{part}}_{jl|i}$ \cite{whittaker1990graphical}, which essentially indicates the surplus correlation between the nodes $j$ and $l$ beyond their indirect correlation through the node $i$. 

In our study we modify the LCC definition from \cite{frontiers2018} by replacing all Pearson's product-moment correlation coefficients by the respective Kendall's rank correlation coefficients, which produces the expression
\begin{equation}\label{eq:LCCw}
LCC_i^{\text{wei}} = \frac{\sum_{1\le j < l \le N, (j,l\neq i)} | \tau_{ij} \cdot \tau_{il} \cdot \tau^{\text{part}}_{jl|i}|}{\sum_{1\le j < l \le N (j,l\neq i)} | \tau_{ij} \cdot \tau_{il}|},
\end{equation}
where the three-way partial Kendall's rank correlation coefficient $\tau^{\text{part}}_{jl|i}$ is defined according to \cite[Section~12.6]{gibbons2003nonparametric} as
\begin{equation}\label{eq:taupart}
\tau^{\text{part}}_{jl|i} = \frac{\tau_{jl} - \tau_{ij} \cdot \tau_{il} }{\sqrt{1 - \tau_{ij}^2} \sqrt{1 - \tau_{il}^2}}.
\end{equation}
Notably, this expression is analogous to the definition of the three-way partial Pearson's correlation coefficient $\rho^{\text{part}}_{jl|i}$ \cite{frontiers2018, whittaker1990graphical} with all pairwise Pearson's correlations replaced by the respective Kendall's correlations.

As long as the expression \eqref{eq:taupart} contains a division by zero whenever $|\tau_{ij}=1|$ or $|\tau_{il}=1|$, such values of the indices $j$ and $l$ are excluded from summation both in the numerator and in the denominator of \eqref{eq:LCCw}, whenever this occurs (such cases turn out to be rare but not impossible in our actual computations). Essentially, as mentioned in Sec.~\ref{sec:kendall}, unity absolute value of Kendall's $\tau$ implies that the bivariate statistics within such a pair of variables (between nodes $i$ and $j$, or $i$ and $l$) is indistinguishable by available data from a deterministic monotone functional dependence. This may be seen as a justification for our decision to exclude such pairs of nodes from summation, which may be interpreted as them being temporarily lumped together into a single node (for a particular position of the time window), as soon as they turn out to behave as a single node anyway. That said, excluding unity-correlated pairs of nodes inevitably introduces a methodical perturbation into the obtained LCC value, and the final justification for this operation relies mostly on our observation that such perturbations, being rare and relatively small, do not affect the capability of thus computed LCC as a climate marker.

As it is pointed out in \cite{frontiers2018}, the use of three-way partial correlations, which naturally appear in \eqref{eq:LCCw} due to the triangle-based nature of the LCC definition, allows to avoid the computation of full partial correlation matrices (taking account of all indirect correlation paths between each pair of nodes), which is a difficult task in terms of both computational burden and poor tractability for statistical estimation \cite{brier2015partial}.

Note also that due to the absolute value taken in \eqref{eq:LCCw}, positive and negative correlations are assumed to contribute equally to the LCC value \cite{frontiers2018}.

It was shown in \cite{frontiers2018} that LCC defined in terms of three-way partial correlations better reveals the age dependence of brain connectivity by functional magnetic resonance data as compared to the LCC definition \eqref{eq:LCC} for unweighted networks (constructed by thresholding the correlation matrix with two threshold values corresponding to fixed edge densities 10\% and 20\%), and as well to other available LCC definitions for weighted networks \cite{saramaki2007generalizations, rubinov2010complex, rubinov2011weight, wang2017comparison}. The present study is the first to apply the LCC definition \eqref{eq:LCCw} to climate networks.

\section{Results}
We have computed snapshots of LCC (namely, its respective formulations \eqref{eq:LCC} or \eqref{eq:LCCw}) for each position of the sliding time window using both the full weighted correlation network and the unweighted network with edge density 5\% obtained by thresholding the correlation matrix. We focus on the relation of the computed LCC value (both versions thereof) to the available data on tropical cyclones in the region of interest, as per the Best Track database \cite{mohapatra2012best}. We represent the results in the form of a color geographic map for LCC (sea area only) for each given position of the time window, superimposed with a track of a cyclone if the current time of observation (defined as the ending time of the time window used to construct the network) falls within the existence period of this cyclone.

\begin{figure*}
	\centering
	\includegraphics[width=0.49\textwidth]{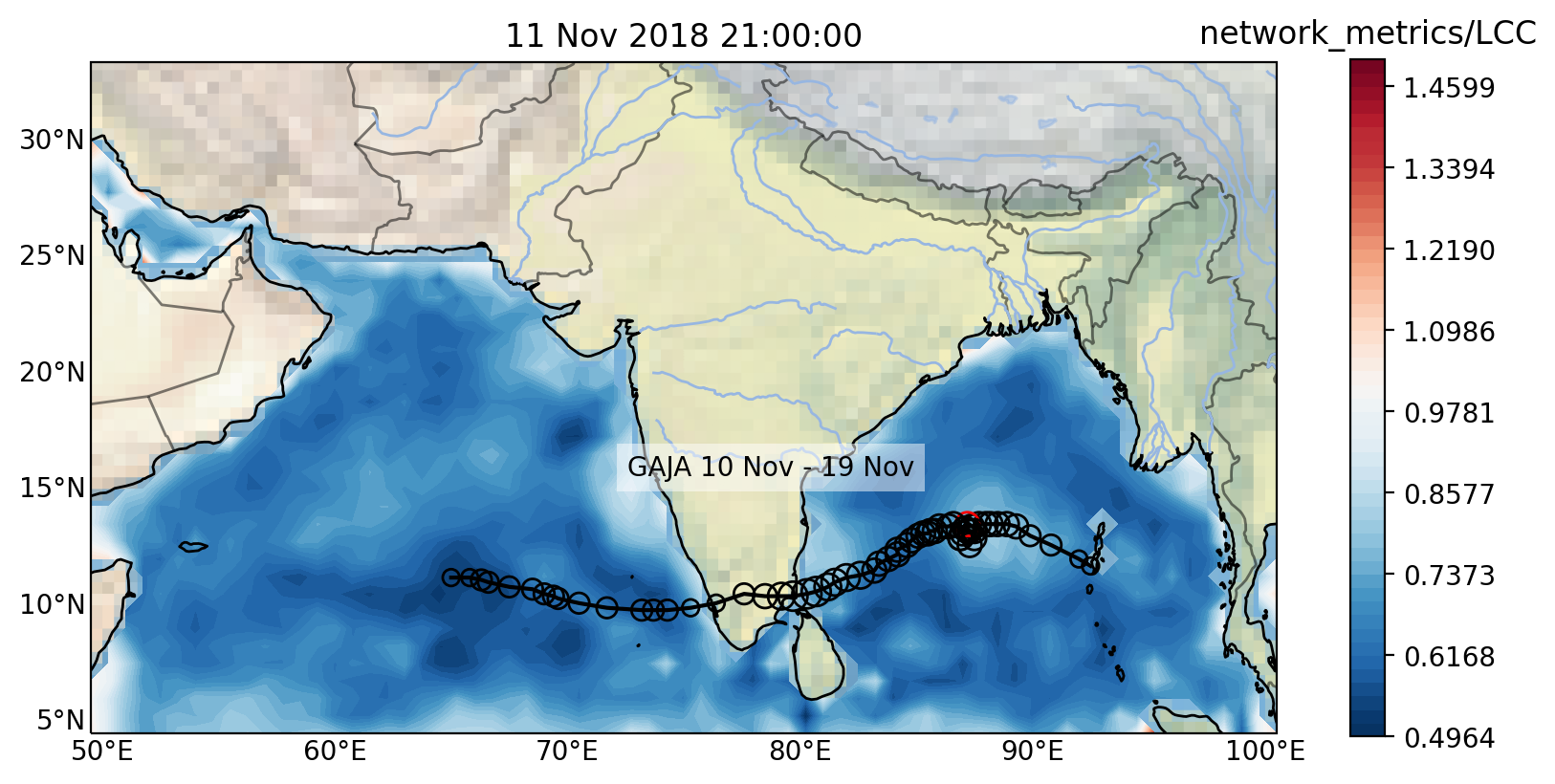}
	\includegraphics[width=0.49\textwidth]{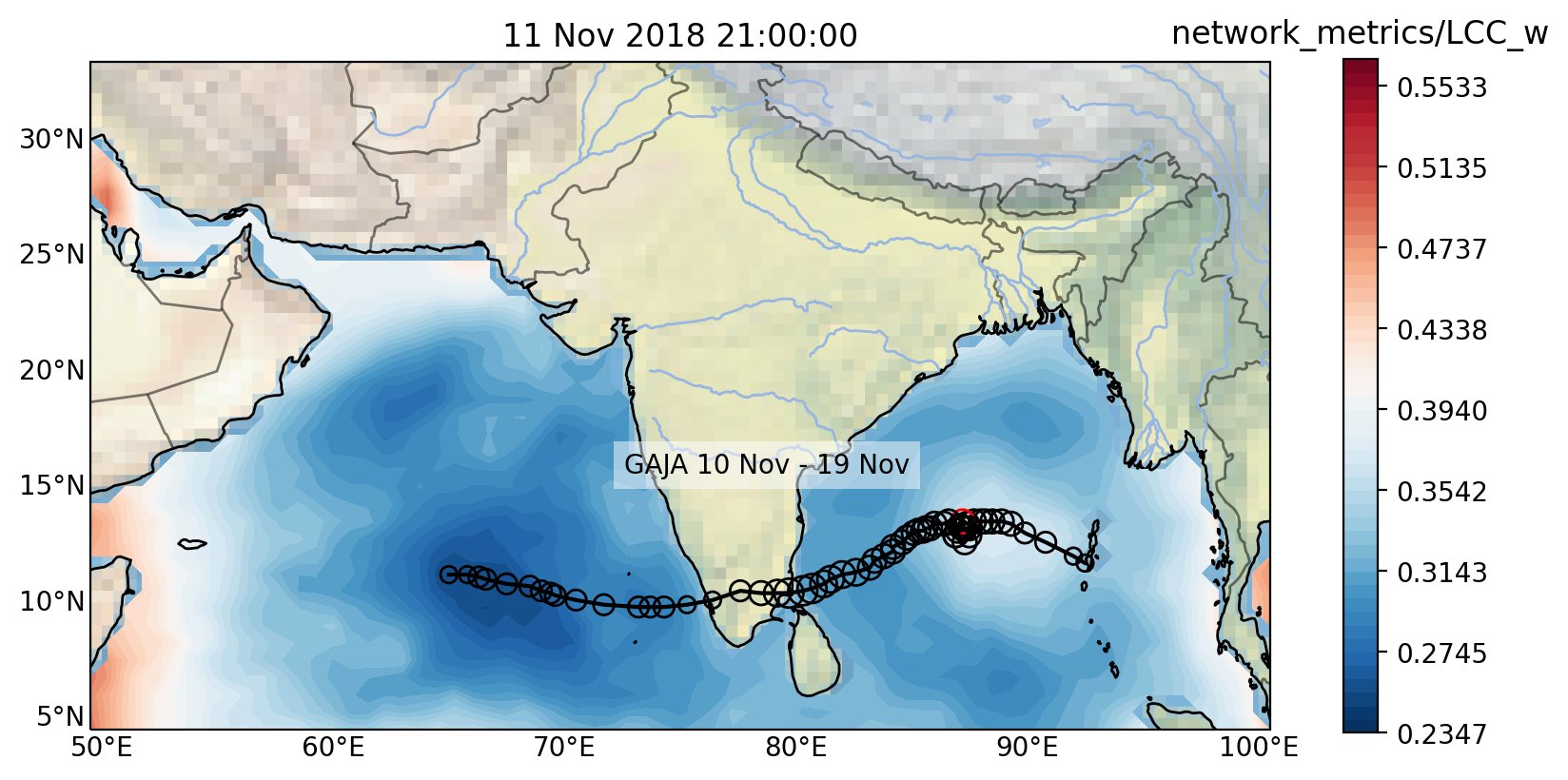}\\
	\includegraphics[width=0.49\textwidth]{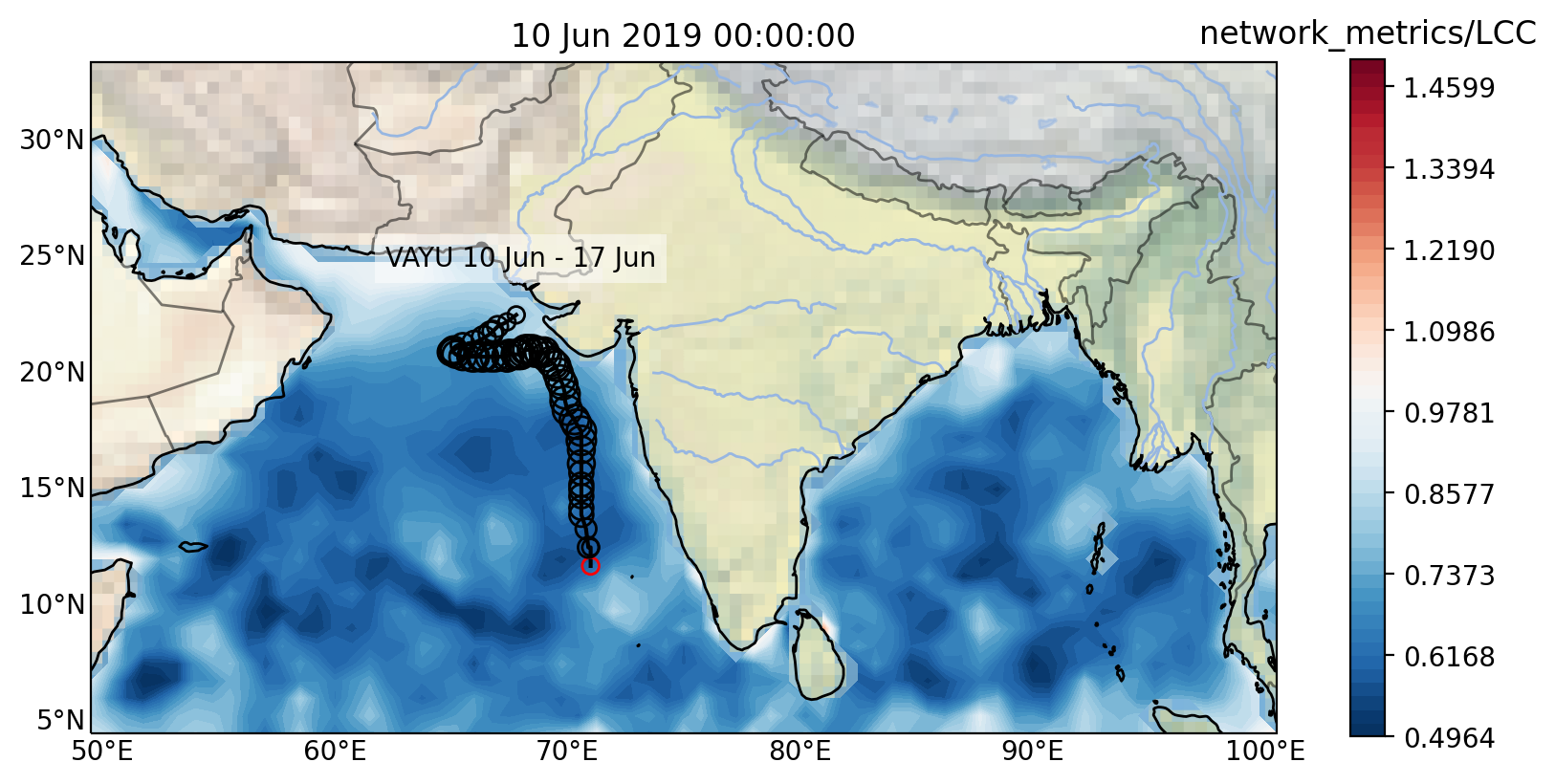}
	\includegraphics[width=0.49\textwidth]{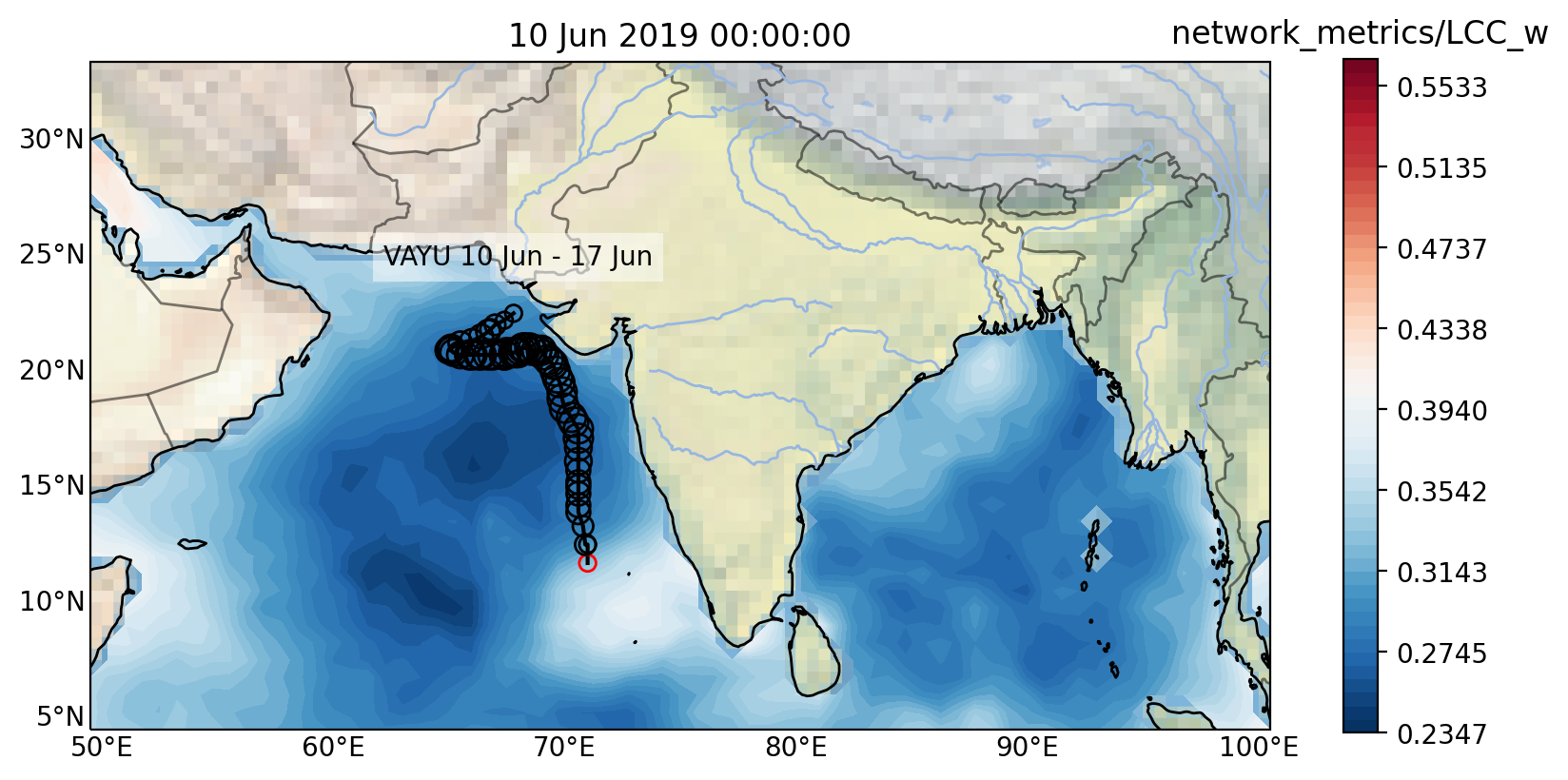}\\
	\includegraphics[width=0.49\textwidth]{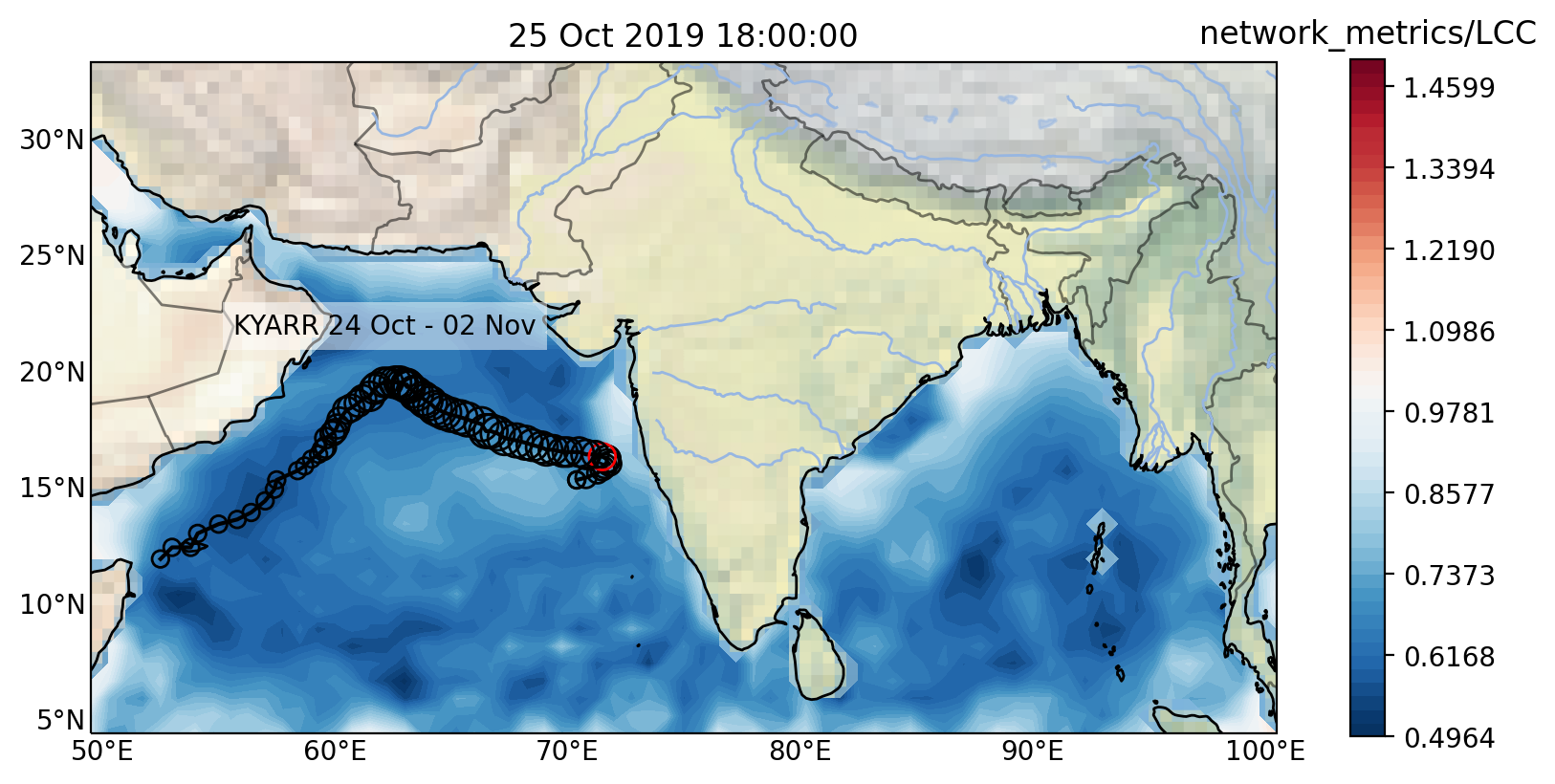}
	\includegraphics[width=0.49\textwidth]{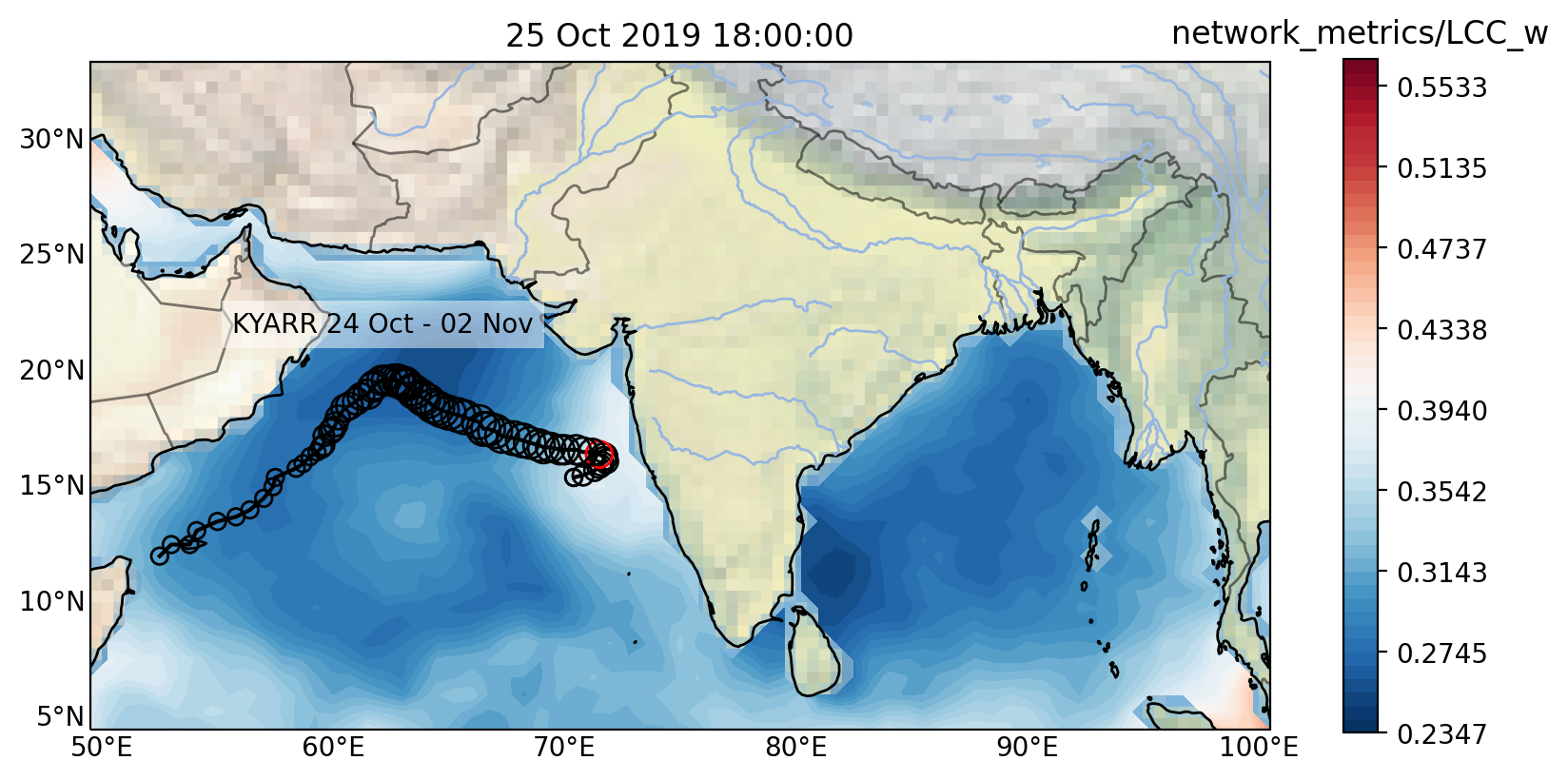}\\
	\includegraphics[width=0.49\textwidth]{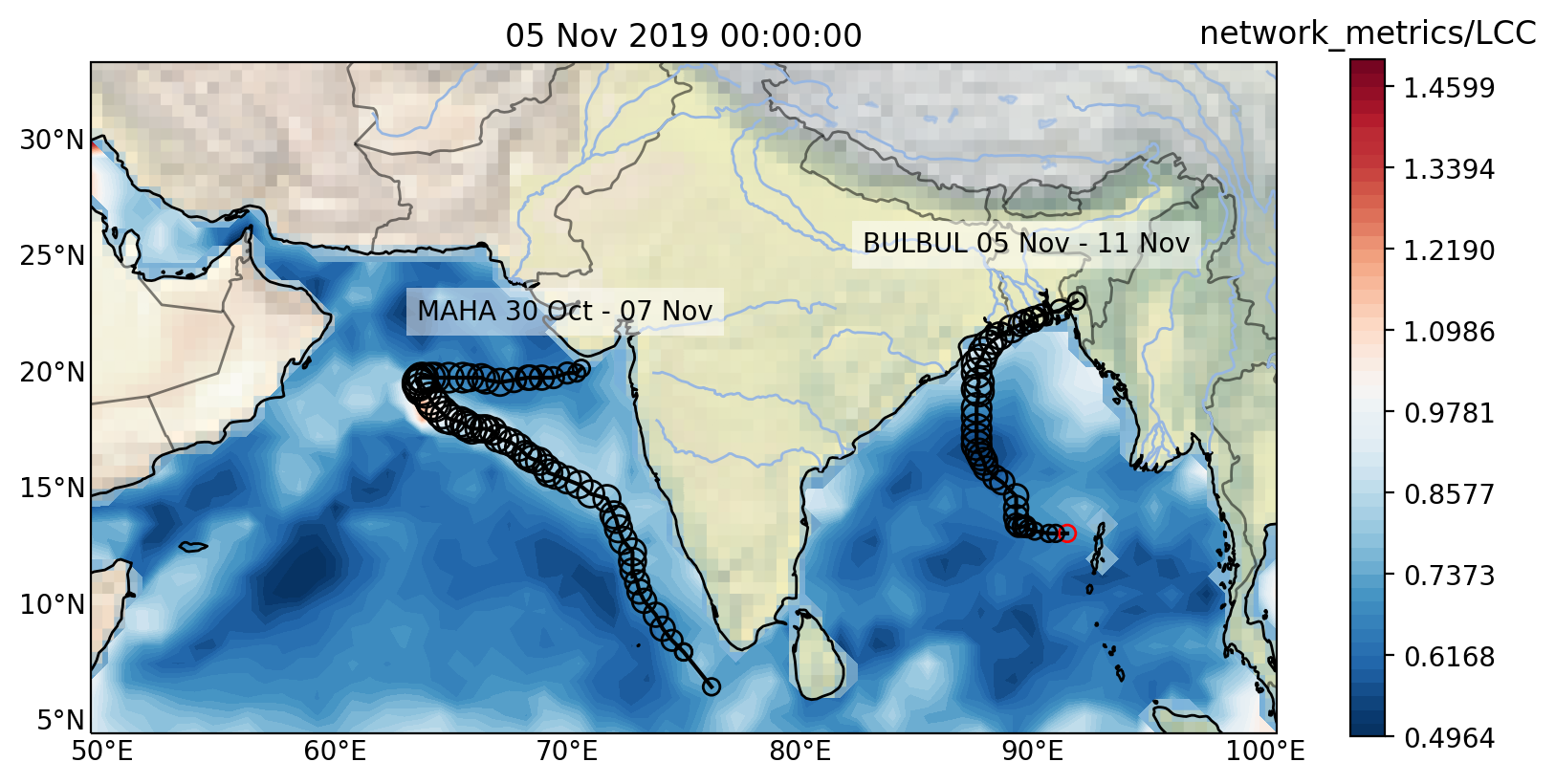}
	\includegraphics[width=0.49\textwidth]{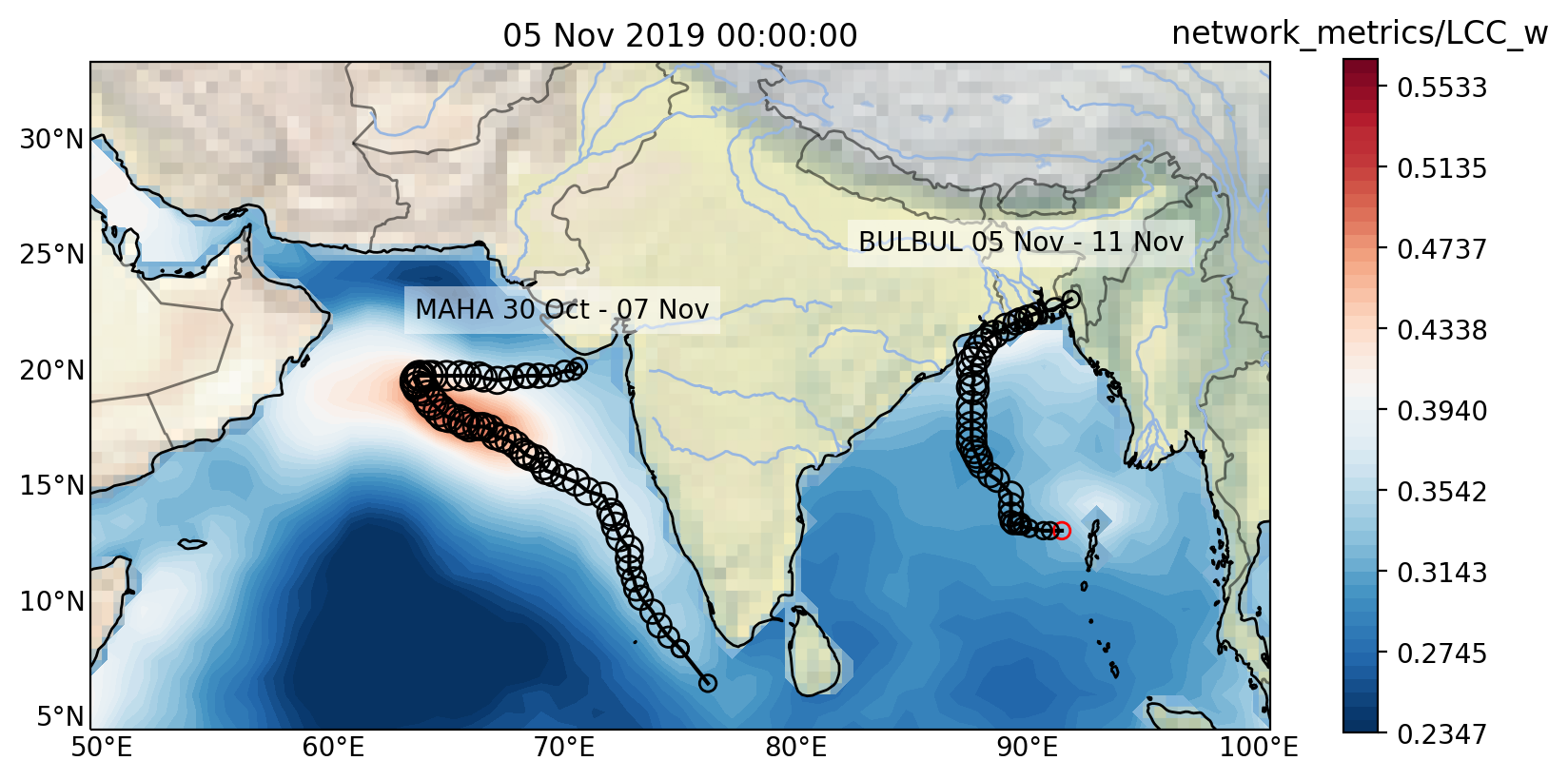}\\
	\caption{Color maps of local clustering coefficient calculated from unweighted (left) and weighted (right) correlation networks of mean sea level pressure anomaly during several selected tropical cyclones in the northern Indian Ocean. Each row corresponds to a particular position of the sliding time window used to construct the networks. Observation time (i.e. the ending time of the window) is indicated above the map. Tracks of tropical cyclones are marked with a series of circles, whose radius denotes the cyclone strength. Name and time period of each cyclone is indicated above the track. Current position of each cyclone at the time of observation is marked with a red circle.}\label{fig:Climate}
\end{figure*}

We aim at demonstrating that the LCC formulation \eqref{eq:LCCw} based on a weighted network may deliver information associated with a tropical cyclone above and beyond the information delivered by the unweighted LCC formulation \eqref{eq:LCC}. In the present study we limit ourselves to substantiating this statement by examples where it is evident by a visual side-by-side comparison of the respective color maps. Several such examples are presented in the Figure~\ref{fig:Climate}, where each row corresponds to a particular moment of observation (indicated above each map) during a tropical cyclone (the name and the time period of each cyclone is indicated above the cyclone track),  color encodes the unweighted (left column) and weighted (right column) versions of LCC at the same observation time. Each cyclone track is marked with a series of circles, whose radius visualizes the cyclone strength in the respective location. The current position of a cyclone (at the time of observation) is marked with a red circle.

The comparison of the respective maps in the Figure~\ref{fig:Climate} on the left and on the right leads to an observation that the weighted version of LCC (on the right) produces a more pronounced local peak in the vicinity of the current location of the cyclone than the unweighted version thereof (on the left). Note that the Figure~\ref{fig:Climate} intentionally shows only examples supporting the statement above. In other cases the weighted LCC version may show little or no improvement over the unweighted version. Nevertheless, when addressing a question, whether a new quantitative indicator (here, the weighted-network LCC formulation) should be included in the toolset of climate network analysis to get the most information of it, even an individual positive example is sufficient to draw a positive conclusion.

\section{Conclusion}
We have found several particular cases where the new proposed network marker defined by Eqs.~\eqref{eq:LCCw} and \eqref{eq:taupart} turns out to be better associated to tropical cyclones than the unweighted local clustering coefficient. This finding alone is sufficient to recommend the usage of this marker among the toolset of climate network analysis, in particular, in application to predicting extreme climate events. Up to this point we have not identified any specific predictive criterion (e.g. for tropical cyclones) based on the proposed marker. Formulating such a criterion, which remains a matter of further study, would additionally make possible a quantitative comparison of the considered network measures in terms of prediction accuracy.

\bibliography{report}

\begin{thebibliography}{34}%
\makeatletter
\providecommand \@ifxundefined [1]{%
 \@ifx{#1\undefined}
}%
\providecommand \@ifnum [1]{%
 \ifnum #1\expandafter \@firstoftwo
 \else \expandafter \@secondoftwo
 \fi
}%
\providecommand \@ifx [1]{%
 \ifx #1\expandafter \@firstoftwo
 \else \expandafter \@secondoftwo
 \fi
}%
\providecommand \natexlab [1]{#1}%
\providecommand \enquote  [1]{``#1''}%
\providecommand \bibnamefont  [1]{#1}%
\providecommand \bibfnamefont [1]{#1}%
\providecommand \citenamefont [1]{#1}%
\providecommand \href@noop [0]{\@secondoftwo}%
\providecommand \href [0]{\begingroup \@sanitize@url \@href}%
\providecommand \@href[1]{\@@startlink{#1}\@@href}%
\providecommand \@@href[1]{\endgroup#1\@@endlink}%
\providecommand \@sanitize@url [0]{\catcode `\\12\catcode `\$12\catcode
  `\&12\catcode `\#12\catcode `\^12\catcode `\_12\catcode `\%12\relax}%
\providecommand \@@startlink[1]{}%
\providecommand \@@endlink[0]{}%
\providecommand \url  [0]{\begingroup\@sanitize@url \@url }%
\providecommand \@url [1]{\endgroup\@href {#1}{\urlprefix }}%
\providecommand \urlprefix  [0]{URL }%
\providecommand \Eprint [0]{\href }%
\providecommand \doibase [0]{http://dx.doi.org/}%
\providecommand \selectlanguage [0]{\@gobble}%
\providecommand \bibinfo  [0]{\@secondoftwo}%
\providecommand \bibfield  [0]{\@secondoftwo}%
\providecommand \translation [1]{[#1]}%
\providecommand \BibitemOpen [0]{}%
\providecommand \bibitemStop [0]{}%
\providecommand \bibitemNoStop [0]{.\EOS\space}%
\providecommand \EOS [0]{\spacefactor3000\relax}%
\providecommand \BibitemShut  [1]{\csname bibitem#1\endcsname}%
\let\auto@bib@innerbib\@empty
\bibitem [{\citenamefont {Langfelder}\ and\ \citenamefont
  {Horvath}(2008)}]{langfelder2008wgcna}%
  \BibitemOpen
  \bibfield  {author} {\bibinfo {author} {\bibfnamefont {P.}~\bibnamefont
  {Langfelder}}\ and\ \bibinfo {author} {\bibfnamefont {S.}~\bibnamefont
  {Horvath}},\ }\href@noop {} {\bibfield  {journal} {\bibinfo  {journal} {BMC
  bioinformatics}\ }\textbf {\bibinfo {volume} {9}},\ \bibinfo {pages} {1}
  (\bibinfo {year} {2008})}\BibitemShut {NoStop}%
\bibitem [{\citenamefont {Bruggeman}\ and\ \citenamefont
  {Westerhoff}(2007)}]{bruggeman2007nature}%
  \BibitemOpen
  \bibfield  {author} {\bibinfo {author} {\bibfnamefont {F.~J.}\ \bibnamefont
  {Bruggeman}}\ and\ \bibinfo {author} {\bibfnamefont {H.~V.}\ \bibnamefont
  {Westerhoff}},\ }\href@noop {} {\bibfield  {journal} {\bibinfo  {journal}
  {TRENDS in Microbiology}\ }\textbf {\bibinfo {volume} {15}},\ \bibinfo
  {pages} {45} (\bibinfo {year} {2007})}\BibitemShut {NoStop}%
\bibitem [{\citenamefont {Van~der Greef}\ \emph {et~al.}(2007)\citenamefont
  {Van~der Greef}, \citenamefont {Martin}, \citenamefont {Juhasz},
  \citenamefont {Adourian}, \citenamefont {Plasterer}, \citenamefont
  {Verheij},\ and\ \citenamefont {McBurney}}]{van2007art}%
  \BibitemOpen
  \bibfield  {author} {\bibinfo {author} {\bibfnamefont {J.}~\bibnamefont
  {Van~der Greef}}, \bibinfo {author} {\bibfnamefont {S.}~\bibnamefont
  {Martin}}, \bibinfo {author} {\bibfnamefont {P.}~\bibnamefont {Juhasz}},
  \bibinfo {author} {\bibfnamefont {A.}~\bibnamefont {Adourian}}, \bibinfo
  {author} {\bibfnamefont {T.}~\bibnamefont {Plasterer}}, \bibinfo {author}
  {\bibfnamefont {E.}~\bibnamefont {Verheij}}, \ and\ \bibinfo {author}
  {\bibfnamefont {R.}~\bibnamefont {McBurney}},\ }\href@noop {} {\bibfield
  {journal} {\bibinfo  {journal} {Journal of proteome research}\ }\textbf
  {\bibinfo {volume} {6}},\ \bibinfo {pages} {1540} (\bibinfo {year}
  {2007})}\BibitemShut {NoStop}%
\bibitem [{\citenamefont {Friedman}\ and\ \citenamefont
  {Alm}(2012)}]{friedman2012inferring}%
  \BibitemOpen
  \bibfield  {author} {\bibinfo {author} {\bibfnamefont {J.}~\bibnamefont
  {Friedman}}\ and\ \bibinfo {author} {\bibfnamefont {E.~J.}\ \bibnamefont
  {Alm}},\ }\href@noop {} {\bibfield  {journal} {\bibinfo  {journal} {PLoS
  Comput Biol}\ }\textbf {\bibinfo {volume} {8}},\ \bibinfo {pages} {e1002687}
  (\bibinfo {year} {2012})}\BibitemShut {NoStop}%
\bibitem [{\citenamefont {Rosato}\ \emph {et~al.}(2018)\citenamefont {Rosato},
  \citenamefont {Tenori}, \citenamefont {Cascante}, \citenamefont {Carulla},
  \citenamefont {Dos~Santos},\ and\ \citenamefont
  {Saccenti}}]{rosato2018correlation}%
  \BibitemOpen
  \bibfield  {author} {\bibinfo {author} {\bibfnamefont {A.}~\bibnamefont
  {Rosato}}, \bibinfo {author} {\bibfnamefont {L.}~\bibnamefont {Tenori}},
  \bibinfo {author} {\bibfnamefont {M.}~\bibnamefont {Cascante}}, \bibinfo
  {author} {\bibfnamefont {P.~R. D.~A.}\ \bibnamefont {Carulla}}, \bibinfo
  {author} {\bibfnamefont {V.~A.~M.}\ \bibnamefont {Dos~Santos}}, \ and\
  \bibinfo {author} {\bibfnamefont {E.}~\bibnamefont {Saccenti}},\ }\href@noop
  {} {\bibfield  {journal} {\bibinfo  {journal} {Metabolomics}\ }\textbf
  {\bibinfo {volume} {14}},\ \bibinfo {pages} {1} (\bibinfo {year}
  {2018})}\BibitemShut {NoStop}%
\bibitem [{\citenamefont {Fransson}\ and\ \citenamefont
  {Marrelec}(2008)}]{fransson2008precuneus}%
  \BibitemOpen
  \bibfield  {author} {\bibinfo {author} {\bibfnamefont {P.}~\bibnamefont
  {Fransson}}\ and\ \bibinfo {author} {\bibfnamefont {G.}~\bibnamefont
  {Marrelec}},\ }\href@noop {} {\bibfield  {journal} {\bibinfo  {journal}
  {Neuroimage}\ }\textbf {\bibinfo {volume} {42}},\ \bibinfo {pages} {1178}
  (\bibinfo {year} {2008})}\BibitemShut {NoStop}%
\bibitem [{\citenamefont {Bassett}\ and\ \citenamefont
  {Sporns}(2017)}]{bassett2017network}%
  \BibitemOpen
  \bibfield  {author} {\bibinfo {author} {\bibfnamefont {D.~S.}\ \bibnamefont
  {Bassett}}\ and\ \bibinfo {author} {\bibfnamefont {O.}~\bibnamefont
  {Sporns}},\ }\href@noop {} {\bibfield  {journal} {\bibinfo  {journal} {Nature
  neuroscience}\ }\textbf {\bibinfo {volume} {20}},\ \bibinfo {pages} {353}
  (\bibinfo {year} {2017})}\BibitemShut {NoStop}%
\bibitem [{\citenamefont {Tsonis}\ \emph {et~al.}(2006)\citenamefont {Tsonis},
  \citenamefont {Swanson},\ and\ \citenamefont {Roebber}}]{tsonis2006networks}%
  \BibitemOpen
  \bibfield  {author} {\bibinfo {author} {\bibfnamefont {A.~A.}\ \bibnamefont
  {Tsonis}}, \bibinfo {author} {\bibfnamefont {K.~L.}\ \bibnamefont {Swanson}},
  \ and\ \bibinfo {author} {\bibfnamefont {P.~J.}\ \bibnamefont {Roebber}},\
  }\href@noop {} {\bibfield  {journal} {\bibinfo  {journal} {Bulletin of the
  American Meteorological Society}\ }\textbf {\bibinfo {volume} {87}},\
  \bibinfo {pages} {585} (\bibinfo {year} {2006})}\BibitemShut {NoStop}%
\bibitem [{\citenamefont {Donges}\ \emph {et~al.}(2009)\citenamefont {Donges},
  \citenamefont {Zou}, \citenamefont {Marwan},\ and\ \citenamefont
  {Kurths}}]{donges2009complex}%
  \BibitemOpen
  \bibfield  {author} {\bibinfo {author} {\bibfnamefont {J.~F.}\ \bibnamefont
  {Donges}}, \bibinfo {author} {\bibfnamefont {Y.}~\bibnamefont {Zou}},
  \bibinfo {author} {\bibfnamefont {N.}~\bibnamefont {Marwan}}, \ and\ \bibinfo
  {author} {\bibfnamefont {J.}~\bibnamefont {Kurths}},\ }\href@noop {}
  {\bibfield  {journal} {\bibinfo  {journal} {The European Physical Journal
  Special Topics}\ }\textbf {\bibinfo {volume} {174}},\ \bibinfo {pages} {157}
  (\bibinfo {year} {2009})}\BibitemShut {NoStop}%
\bibitem [{\citenamefont {Gozolchiani}\ \emph {et~al.}(2011)\citenamefont
  {Gozolchiani}, \citenamefont {Havlin},\ and\ \citenamefont
  {Yamasaki}}]{gozolchiani2011emergence}%
  \BibitemOpen
  \bibfield  {author} {\bibinfo {author} {\bibfnamefont {A.}~\bibnamefont
  {Gozolchiani}}, \bibinfo {author} {\bibfnamefont {S.}~\bibnamefont {Havlin}},
  \ and\ \bibinfo {author} {\bibfnamefont {K.}~\bibnamefont {Yamasaki}},\
  }\href@noop {} {\bibfield  {journal} {\bibinfo  {journal} {Physical review
  letters}\ }\textbf {\bibinfo {volume} {107}},\ \bibinfo {pages} {148501}
  (\bibinfo {year} {2011})}\BibitemShut {NoStop}%
\bibitem [{\citenamefont {Dijkstra}\ \emph
  {et~al.}(2019{\natexlab{a}})\citenamefont {Dijkstra}, \citenamefont
  {Hern{\'a}ndez-Garc{\'\i}a}, \citenamefont {Masoller},\ and\ \citenamefont
  {Barreiro}}]{dijkstra2019networks}%
  \BibitemOpen
  \bibfield  {author} {\bibinfo {author} {\bibfnamefont {H.~A.}\ \bibnamefont
  {Dijkstra}}, \bibinfo {author} {\bibfnamefont {E.}~\bibnamefont
  {Hern{\'a}ndez-Garc{\'\i}a}}, \bibinfo {author} {\bibfnamefont
  {C.}~\bibnamefont {Masoller}}, \ and\ \bibinfo {author} {\bibfnamefont
  {M.}~\bibnamefont {Barreiro}},\ }\href@noop {} {\emph {\bibinfo {title}
  {Networks in Climate}}}\ (\bibinfo  {publisher} {Cambridge University
  Press},\ \bibinfo {year} {2019})\BibitemShut {NoStop}%
\bibitem [{\citenamefont {Feng}\ \emph {et~al.}(2016)\citenamefont {Feng},
  \citenamefont {Vasile}, \citenamefont {Segond}, \citenamefont {Gozolchiani},
  \citenamefont {Wang}, \citenamefont {Abel}, \citenamefont {Havlin},
  \citenamefont {Bunde},\ and\ \citenamefont
  {Dijkstra}}]{feng2016climatelearn}%
  \BibitemOpen
  \bibfield  {author} {\bibinfo {author} {\bibfnamefont {Q.~Y.}\ \bibnamefont
  {Feng}}, \bibinfo {author} {\bibfnamefont {R.}~\bibnamefont {Vasile}},
  \bibinfo {author} {\bibfnamefont {M.}~\bibnamefont {Segond}}, \bibinfo
  {author} {\bibfnamefont {A.}~\bibnamefont {Gozolchiani}}, \bibinfo {author}
  {\bibfnamefont {Y.}~\bibnamefont {Wang}}, \bibinfo {author} {\bibfnamefont
  {M.}~\bibnamefont {Abel}}, \bibinfo {author} {\bibfnamefont {S.}~\bibnamefont
  {Havlin}}, \bibinfo {author} {\bibfnamefont {A.}~\bibnamefont {Bunde}}, \
  and\ \bibinfo {author} {\bibfnamefont {H.~A.}\ \bibnamefont {Dijkstra}},\
  }\href@noop {} {\bibfield  {journal} {\bibinfo  {journal} {Geoscientific
  Model Development}\ } (\bibinfo {year} {2016})}\BibitemShut {NoStop}%
\bibitem [{\citenamefont {Dijkstra}\ \emph
  {et~al.}(2019{\natexlab{b}})\citenamefont {Dijkstra}, \citenamefont
  {Hernandez-Garcia}, \citenamefont {Lopez} \emph
  {et~al.}}]{dijkstra2019application}%
  \BibitemOpen
  \bibfield  {author} {\bibinfo {author} {\bibfnamefont {H.}~\bibnamefont
  {Dijkstra}}, \bibinfo {author} {\bibfnamefont {E.}~\bibnamefont
  {Hernandez-Garcia}}, \bibinfo {author} {\bibfnamefont {C.}~\bibnamefont
  {Lopez}},  \emph {et~al.},\ }\href@noop {} {\bibfield  {journal} {\bibinfo
  {journal} {Frontiers in Physics}\ }\textbf {\bibinfo {volume} {7}},\ \bibinfo
  {pages} {153} (\bibinfo {year} {2019}{\natexlab{b}})}\BibitemShut {NoStop}%
\bibitem [{\citenamefont {Bockmayr}\ \emph {et~al.}(2013)\citenamefont
  {Bockmayr}, \citenamefont {Klauschen}, \citenamefont {Gy{\"o}rffy},
  \citenamefont {Denkert},\ and\ \citenamefont {Budczies}}]{bockmayr2013new}%
  \BibitemOpen
  \bibfield  {author} {\bibinfo {author} {\bibfnamefont {M.}~\bibnamefont
  {Bockmayr}}, \bibinfo {author} {\bibfnamefont {F.}~\bibnamefont {Klauschen}},
  \bibinfo {author} {\bibfnamefont {B.}~\bibnamefont {Gy{\"o}rffy}}, \bibinfo
  {author} {\bibfnamefont {C.}~\bibnamefont {Denkert}}, \ and\ \bibinfo
  {author} {\bibfnamefont {J.}~\bibnamefont {Budczies}},\ }\href@noop {}
  {\bibfield  {journal} {\bibinfo  {journal} {BMC systems biology}\ }\textbf
  {\bibinfo {volume} {7}},\ \bibinfo {pages} {1} (\bibinfo {year}
  {2013})}\BibitemShut {NoStop}%
\bibitem [{\citenamefont {Newman}(2016)}]{Newman2016}%
  \BibitemOpen
  \bibfield  {author} {\bibinfo {author} {\bibfnamefont {M.~E.~J.}\
  \bibnamefont {Newman}},\ }in\ \href@noop {} {\emph {\bibinfo {booktitle} {The
  New Palgrave Dictionary of Economics}}}\ (\bibinfo  {publisher} {Palgrave
  Macmillan UK},\ \bibinfo {address} {London},\ \bibinfo {year} {2016})\ pp.\
  \bibinfo {pages} {1--8}\BibitemShut {NoStop}%
\bibitem [{\citenamefont {Masuda}\ \emph {et~al.}(2018)\citenamefont {Masuda},
  \citenamefont {Sakaki}, \citenamefont {Ezaki},\ and\ \citenamefont
  {Watanabe}}]{frontiers2018}%
  \BibitemOpen
  \bibfield  {author} {\bibinfo {author} {\bibfnamefont {N.}~\bibnamefont
  {Masuda}}, \bibinfo {author} {\bibfnamefont {M.}~\bibnamefont {Sakaki}},
  \bibinfo {author} {\bibfnamefont {T.}~\bibnamefont {Ezaki}}, \ and\ \bibinfo
  {author} {\bibfnamefont {T.}~\bibnamefont {Watanabe}},\ }\href {\doibase
  10.3389/fninf.2018.00007} {\bibfield  {journal} {\bibinfo  {journal}
  {Frontiers in Neuroinformatics}\ }\textbf {\bibinfo {volume} {12}},\ \bibinfo
  {pages} {7} (\bibinfo {year} {2018})}\BibitemShut {NoStop}%
\bibitem [{\citenamefont {Bonett}\ and\ \citenamefont
  {Wright}(2000)}]{bonett2000sample}%
  \BibitemOpen
  \bibfield  {author} {\bibinfo {author} {\bibfnamefont {D.~G.}\ \bibnamefont
  {Bonett}}\ and\ \bibinfo {author} {\bibfnamefont {T.~A.}\ \bibnamefont
  {Wright}},\ }\href@noop {} {\bibfield  {journal} {\bibinfo  {journal}
  {Psychometrika}\ }\textbf {\bibinfo {volume} {65}},\ \bibinfo {pages} {23}
  (\bibinfo {year} {2000})}\BibitemShut {NoStop}%
\bibitem [{\citenamefont {Hersbach}\ \emph {et~al.}(2020)\citenamefont
  {Hersbach}, \citenamefont {Bell}, \citenamefont {Berrisford}, \citenamefont
  {Hirahara}, \citenamefont {Hor{\'a}nyi}, \citenamefont {Mu{\~n}oz-Sabater},
  \citenamefont {Nicolas}, \citenamefont {Peubey}, \citenamefont {Radu},
  \citenamefont {Schepers} \emph {et~al.}}]{hersbach2020era5}%
  \BibitemOpen
  \bibfield  {author} {\bibinfo {author} {\bibfnamefont {H.}~\bibnamefont
  {Hersbach}}, \bibinfo {author} {\bibfnamefont {B.}~\bibnamefont {Bell}},
  \bibinfo {author} {\bibfnamefont {P.}~\bibnamefont {Berrisford}}, \bibinfo
  {author} {\bibfnamefont {S.}~\bibnamefont {Hirahara}}, \bibinfo {author}
  {\bibfnamefont {A.}~\bibnamefont {Hor{\'a}nyi}}, \bibinfo {author}
  {\bibfnamefont {J.}~\bibnamefont {Mu{\~n}oz-Sabater}}, \bibinfo {author}
  {\bibfnamefont {J.}~\bibnamefont {Nicolas}}, \bibinfo {author} {\bibfnamefont
  {C.}~\bibnamefont {Peubey}}, \bibinfo {author} {\bibfnamefont
  {R.}~\bibnamefont {Radu}}, \bibinfo {author} {\bibfnamefont {D.}~\bibnamefont
  {Schepers}},  \emph {et~al.},\ }\href@noop {} {\bibfield  {journal} {\bibinfo
   {journal} {Quarterly Journal of the Royal Meteorological Society}\ }\textbf
  {\bibinfo {volume} {146}},\ \bibinfo {pages} {1999} (\bibinfo {year}
  {2020})}\BibitemShut {NoStop}%
\bibitem [{\citenamefont {Mohapatra}\ \emph {et~al.}(2012)\citenamefont
  {Mohapatra}, \citenamefont {Bandyopadhyay},\ and\ \citenamefont
  {Tyagi}}]{mohapatra2012best}%
  \BibitemOpen
  \bibfield  {author} {\bibinfo {author} {\bibfnamefont {M.}~\bibnamefont
  {Mohapatra}}, \bibinfo {author} {\bibfnamefont {B.}~\bibnamefont
  {Bandyopadhyay}}, \ and\ \bibinfo {author} {\bibfnamefont {A.}~\bibnamefont
  {Tyagi}},\ }\href@noop {} {\bibfield  {journal} {\bibinfo  {journal} {Natural
  Hazards}\ }\textbf {\bibinfo {volume} {63}},\ \bibinfo {pages} {1285}
  (\bibinfo {year} {2012})}\BibitemShut {NoStop}%
\bibitem [{\citenamefont {Kendall}(1938)}]{kendall1938new}%
  \BibitemOpen
  \bibfield  {author} {\bibinfo {author} {\bibfnamefont {M.~G.}\ \bibnamefont
  {Kendall}},\ }\href@noop {} {\bibfield  {journal} {\bibinfo  {journal}
  {Biometrika}\ }\textbf {\bibinfo {volume} {30}},\ \bibinfo {pages} {81}
  (\bibinfo {year} {1938})}\BibitemShut {NoStop}%
\bibitem [{Note1()}]{Note1}%
  \BibitemOpen
  \bibinfo {note} {A pair of bivariate observations $(x_1(t_i), x_2(t_i))$ and
  $(x_1(t_j), x_2(t_j))$ is called concordant (discordant), if the product
  $(x_1(t_i)-x_1(t_j)) (x_2(t_i)-x_2(t_j))$ is positive (negative), implying
  that this particular pair of observations shows an increasing (decreasing)
  dependence between $x_1$ and $x_2$.}\BibitemShut {Stop}%
\bibitem [{\citenamefont {Amerise}\ and\ \citenamefont
  {Tarsitano}(2015)}]{amerise2015correction}%
  \BibitemOpen
  \bibfield  {author} {\bibinfo {author} {\bibfnamefont {I.~L.}\ \bibnamefont
  {Amerise}}\ and\ \bibinfo {author} {\bibfnamefont {A.}~\bibnamefont
  {Tarsitano}},\ }\href@noop {} {\bibfield  {journal} {\bibinfo  {journal}
  {Journal of Applied Statistics}\ }\textbf {\bibinfo {volume} {42}},\ \bibinfo
  {pages} {2584} (\bibinfo {year} {2015})}\BibitemShut {NoStop}%
\bibitem [{\citenamefont {Adler}(1957)}]{adler1957modification}%
  \BibitemOpen
  \bibfield  {author} {\bibinfo {author} {\bibfnamefont {L.~M.}\ \bibnamefont
  {Adler}},\ }\href@noop {} {\bibfield  {journal} {\bibinfo  {journal} {Journal
  of the American Statistical Association}\ }\textbf {\bibinfo {volume} {52}},\
  \bibinfo {pages} {33} (\bibinfo {year} {1957})}\BibitemShut {NoStop}%
\bibitem [{\citenamefont {Kruskal}(1958)}]{kruskal1958ordinal}%
  \BibitemOpen
  \bibfield  {author} {\bibinfo {author} {\bibfnamefont {W.~H.}\ \bibnamefont
  {Kruskal}},\ }\href@noop {} {\bibfield  {journal} {\bibinfo  {journal}
  {Journal of the American Statistical Association}\ }\textbf {\bibinfo
  {volume} {53}},\ \bibinfo {pages} {814} (\bibinfo {year} {1958})}\BibitemShut
  {NoStop}%
\bibitem [{\citenamefont {Gibbons}\ and\ \citenamefont
  {Chakraborti}(2003)}]{gibbons2003nonparametric}%
  \BibitemOpen
  \bibfield  {author} {\bibinfo {author} {\bibfnamefont {J.}~\bibnamefont
  {Gibbons}}\ and\ \bibinfo {author} {\bibfnamefont {S.}~\bibnamefont
  {Chakraborti}},\ }\href@noop {} {\emph {\bibinfo {title} {Nonparametric
  Statistical Inference: Fourth Edition, Revised and Expanded. Statistics: A
  Series of Textbooks and Monographs}}}\ (\bibinfo  {publisher} {New York, NY:
  Marcel Dekker, Inc.},\ \bibinfo {year} {2003})\BibitemShut {NoStop}%
\bibitem [{\citenamefont {Goodman}\ and\ \citenamefont
  {Kruskal}(1954)}]{goodman1954measures}%
  \BibitemOpen
  \bibfield  {author} {\bibinfo {author} {\bibfnamefont {L.~A.}\ \bibnamefont
  {Goodman}}\ and\ \bibinfo {author} {\bibfnamefont {W.~H.}\ \bibnamefont
  {Kruskal}},\ }\href@noop {} {\bibfield  {journal} {\bibinfo  {journal}
  {Journal of the American Statistical Association}\ }\textbf {\bibinfo
  {volume} {49}},\ \bibinfo {pages} {732} (\bibinfo {year} {1954})}\BibitemShut
  {NoStop}%
\bibitem [{\citenamefont {Xu}\ \emph {et~al.}(2013)\citenamefont {Xu},
  \citenamefont {Hou}, \citenamefont {Hung},\ and\ \citenamefont
  {Zou}}]{xu2013comparative}%
  \BibitemOpen
  \bibfield  {author} {\bibinfo {author} {\bibfnamefont {W.}~\bibnamefont
  {Xu}}, \bibinfo {author} {\bibfnamefont {Y.}~\bibnamefont {Hou}}, \bibinfo
  {author} {\bibfnamefont {Y.}~\bibnamefont {Hung}}, \ and\ \bibinfo {author}
  {\bibfnamefont {Y.}~\bibnamefont {Zou}},\ }\href@noop {} {\bibfield
  {journal} {\bibinfo  {journal} {Signal Processing}\ }\textbf {\bibinfo
  {volume} {93}},\ \bibinfo {pages} {261} (\bibinfo {year} {2013})}\BibitemShut
  {NoStop}%
\bibitem [{\citenamefont {Watts}\ and\ \citenamefont
  {Strogatz}(1998)}]{watts1998collective}%
  \BibitemOpen
  \bibfield  {author} {\bibinfo {author} {\bibfnamefont {D.~J.}\ \bibnamefont
  {Watts}}\ and\ \bibinfo {author} {\bibfnamefont {S.~H.}\ \bibnamefont
  {Strogatz}},\ }\href@noop {} {\bibfield  {journal} {\bibinfo  {journal}
  {Nature}\ }\textbf {\bibinfo {volume} {393}},\ \bibinfo {pages} {440}
  (\bibinfo {year} {1998})}\BibitemShut {NoStop}%
\bibitem [{\citenamefont {Saram{\"a}ki}\ \emph {et~al.}(2007)\citenamefont
  {Saram{\"a}ki}, \citenamefont {Kivel{\"a}}, \citenamefont {Onnela},
  \citenamefont {Kaski},\ and\ \citenamefont
  {Kertesz}}]{saramaki2007generalizations}%
  \BibitemOpen
  \bibfield  {author} {\bibinfo {author} {\bibfnamefont {J.}~\bibnamefont
  {Saram{\"a}ki}}, \bibinfo {author} {\bibfnamefont {M.}~\bibnamefont
  {Kivel{\"a}}}, \bibinfo {author} {\bibfnamefont {J.-P.}\ \bibnamefont
  {Onnela}}, \bibinfo {author} {\bibfnamefont {K.}~\bibnamefont {Kaski}}, \
  and\ \bibinfo {author} {\bibfnamefont {J.}~\bibnamefont {Kertesz}},\
  }\href@noop {} {\bibfield  {journal} {\bibinfo  {journal} {Physical Review
  E}\ }\textbf {\bibinfo {volume} {75}},\ \bibinfo {pages} {027105} (\bibinfo
  {year} {2007})}\BibitemShut {NoStop}%
\bibitem [{\citenamefont {Rubinov}\ and\ \citenamefont
  {Sporns}(2010)}]{rubinov2010complex}%
  \BibitemOpen
  \bibfield  {author} {\bibinfo {author} {\bibfnamefont {M.}~\bibnamefont
  {Rubinov}}\ and\ \bibinfo {author} {\bibfnamefont {O.}~\bibnamefont
  {Sporns}},\ }\href@noop {} {\bibfield  {journal} {\bibinfo  {journal}
  {Neuroimage}\ }\textbf {\bibinfo {volume} {52}},\ \bibinfo {pages} {1059}
  (\bibinfo {year} {2010})}\BibitemShut {NoStop}%
\bibitem [{\citenamefont {Rubinov}\ and\ \citenamefont
  {Sporns}(2011)}]{rubinov2011weight}%
  \BibitemOpen
  \bibfield  {author} {\bibinfo {author} {\bibfnamefont {M.}~\bibnamefont
  {Rubinov}}\ and\ \bibinfo {author} {\bibfnamefont {O.}~\bibnamefont
  {Sporns}},\ }\href@noop {} {\bibfield  {journal} {\bibinfo  {journal}
  {Neuroimage}\ }\textbf {\bibinfo {volume} {56}},\ \bibinfo {pages} {2068}
  (\bibinfo {year} {2011})}\BibitemShut {NoStop}%
\bibitem [{\citenamefont {Wang}\ \emph {et~al.}(2017)\citenamefont {Wang},
  \citenamefont {Ghumare}, \citenamefont {Vandenberghe},\ and\ \citenamefont
  {Dupont}}]{wang2017comparison}%
  \BibitemOpen
  \bibfield  {author} {\bibinfo {author} {\bibfnamefont {Y.}~\bibnamefont
  {Wang}}, \bibinfo {author} {\bibfnamefont {E.}~\bibnamefont {Ghumare}},
  \bibinfo {author} {\bibfnamefont {R.}~\bibnamefont {Vandenberghe}}, \ and\
  \bibinfo {author} {\bibfnamefont {P.}~\bibnamefont {Dupont}},\ }\href@noop {}
  {\bibfield  {journal} {\bibinfo  {journal} {Neural computation}\ }\textbf
  {\bibinfo {volume} {29}},\ \bibinfo {pages} {313} (\bibinfo {year}
  {2017})}\BibitemShut {NoStop}%
\bibitem [{\citenamefont {Whittaker}\ and\ \citenamefont
  {Whittaker}(1990)}]{whittaker1990graphical}%
  \BibitemOpen
  \bibfield  {author} {\bibinfo {author} {\bibfnamefont {J.}~\bibnamefont
  {Whittaker}}\ and\ \bibinfo {author} {\bibfnamefont {J.}~\bibnamefont
  {Whittaker}},\ }\href@noop {} {\emph {\bibinfo {title} {Graphical models in
  applied multivariate statistics}}},\ Vol.~\bibinfo {volume} {19}\ (\bibinfo
  {publisher} {Wiley Chichester},\ \bibinfo {year} {1990})\BibitemShut
  {NoStop}%
\bibitem [{\citenamefont {Brier}\ \emph {et~al.}(2015)\citenamefont {Brier},
  \citenamefont {Mitra}, \citenamefont {McCarthy}, \citenamefont {Ances},\ and\
  \citenamefont {Snyder}}]{brier2015partial}%
  \BibitemOpen
  \bibfield  {author} {\bibinfo {author} {\bibfnamefont {M.~R.}\ \bibnamefont
  {Brier}}, \bibinfo {author} {\bibfnamefont {A.}~\bibnamefont {Mitra}},
  \bibinfo {author} {\bibfnamefont {J.~E.}\ \bibnamefont {McCarthy}}, \bibinfo
  {author} {\bibfnamefont {B.~M.}\ \bibnamefont {Ances}}, \ and\ \bibinfo
  {author} {\bibfnamefont {A.~Z.}\ \bibnamefont {Snyder}},\ }\href@noop {}
  {\bibfield  {journal} {\bibinfo  {journal} {NeuroImage}\ }\textbf {\bibinfo
  {volume} {121}},\ \bibinfo {pages} {29} (\bibinfo {year} {2015})}\BibitemShut
  {NoStop}%
\end{thebibliography}%

\end{document}